\documentstyle[11pt]{article}
\newcommand{\cita}[1]{\cite{#1}} 
\def\sign(#1){(\!-\!1)^{#1}}
\def\binom(#1,#2){ (\!\!
     \begin{array}{c} #1 \\ #2 \end{array}\!\! ) }
\def\plus{\!+\!}
\def\minus{\!-\!}

\def\nn{\nonumber \\ &&}

\def\Li{\hbox{Li}}
\def\Hpl{\hbox{H}}

\begin{document}
\topskip 2cm
\hfill NIKHEF-99-005

\hfill TTP99-08

\begin{center} {\Large\bf Harmonic Polylogarithms } 
\end{center} 
\vspace{1.2cm} 
\begin{center} 
{ {\large E.~Remiddi$^{a,b}$ } and {\large J. A. M. Vermaseren$^{c}$ \\ } } 
\end{center} 
\begin{itemize} 
\item[$^a$] { \sl Institut f\"ur Theoretische TeilchenPhysik \\
			University of Karlsruhe - D76128 Karlsruhe, Germany \\
			Supported by the Alexander von Humboldt Stiftung }
\item[$^b$] {\sl Dipartimento di Fisica, Universit\`a di Bologna, 
                 I-40126 Bologna, Italy } \\
			{\sl INFN, Sezione di Bologna, I-40126 Bologna, Italy } 
\item[$^c$] {\sl NIKHEF, PO Box 41882, 1009DB Amsterdam, The Netherlands } 
\end{itemize}
\begin{center} 
e-mail: \\
{\tt remiddi@bo.infn.it \\
t68@nikhef.nl \\ }
\end{center} 
\begin{center}
\begin{abstract} 
The harmonic polylogarithms (hpl's) are introduced. They are a 
generalization of Nielsen's polylogarithms, satisfying a product algebra 
(the product of two hpl's is in turn a combination of hpl's) 
and forming a set closed under the transformation of the arguments 
\( x=1/z \) and \( x=(1-t)/(1+t) \). The coefficients of their expansions 
and their Mellin transforms are harmonic sums.
\end{abstract}
\end{center} 

\noindent AMS(1991) subject classification: 
Primary 33B99, 33E99, Secondary 11Y60 
\vspace{1cm} 
\pagestyle{plain} \pagenumbering{arabic} 
\def\a{\alpha} 
\def\app{{\left(\frac{\alpha}{\pi}\right)}} 
\newcommand{\Eq}[1]{Eq.(\ref{#1})} 
\newcommand{\labbel}[1]{\label{#1}} 
\newcommand{\dnk}[1]{ \frac{d^nk_{#1}}{(2\pi)^{n-2}} } 
\newcommand{\e}{{\mathrm{e}}} 
\newcommand{\om}{\omega} 
\newcommand{\ieps}{i\epsilon} 
\newcommand{\verso}[1]{ {\; \buildrel {x \to #1} \over{\longrightarrow}}\; } 
\newcommand{\G}{{{\rm G}}}
\newcommand{\intG}{-\int_0^1 \frac{dt}{t-1/x}} 

\section{ Introduction. } 

Euler's dilogarithm appeared very soon, even if with a different 
name, in the evaluation of radiative corrections in QED. The first 
occurrence is perhaps in the 1934 paper by G. Racah on the radiation 
by fast particles \cita{Racah}, whose function $F(x)$ is equal to 
$-\Li_2(-x)$ in Euler's notation. Two loop calculations \cita{KMR} 
required the polylogarithms, Nielsen's generalization \cita{Nielsen} 
of Euler's dilogarithm. More bibliographical indications as well as 
many relevant results are contained in the popular book by Lewin 
\cita{Lewin} (note the change in the titles of the two editions of the 
book). \\ 
While the polylogarithms are the natural analytical tool to use when 
dealing with the (relatively) simple integrals appearing in calculations 
with a few loops, it is known that they will not be sufficient 
when the number of loops will be larger than has been considered thus far 
or when several different scales are present. 
In a recent publication the set of polylogarithms has been extended into 
something called `multidimensional polylogarithms'~\cita{BBBL}. These 
functions seem to be very useful when more than one dimensionful parameter 
is involved. In principle they are a direct generalization of the 
definition of the power-series expansion of the polylogarithms to a 
multiparameter space.

Besides the dilogarithm, Euler studied also harmonic sums. A recent 
publication by one of us~\cita{HS} investigated harmonic sums and their 
applicability, in particular to formulas in Mellin space. These harmonic 
sums seem to be the natural functions for the results of moment 
calculations of deep inelastic structure functions when only massless 
quarks are involved\footnote{This can be shown for all two loop 
calculations to any order in the expansion parameter $\epsilon$. For three 
loop calculations such results do not exist yet, but a recent result by 
Broadhurst and Kreimer~\cite{BK} shows that only at the 7-loop level the 
counter terms in the QCD beta function contains non-zeta like constants.}. 
If indeed all these moments can be expressed in terms of harmonic sums, the 
class of functions that will represent the results in the regular $x$-space 
will be formed by the inverse Mellin transforms of these harmonic sums. In 
ref~\cita{HS} it was indicated how one could obtain at least numerical 
representations of these functions by means of numerical integration.

In the current paper we study these functions in a more systematic way. We 
start with a recursive integral definition of a class of functions, which 
we will call the harmonic polylogarithms (hpl's), which are by construction 
a generalization of Nielsen's polylogarithms; it turns out, 
further, that an important subset of the hpl's is also a subset of the 
multidimensional polylogarithms of ref~\cita{BBBL}. Then we will study a 
number of their properties, including expressions for products of harmonic 
polylogarithms with the same argument, the behaviour at $x=0,1$, the 
relevant expansions around those points, the algebra of the hpl's and 
the identities between hpl's of related arguments. Then we study special 
values and numerical evaluation.
Finally we study the Mellin transforms of the harmonic polylogarithms and 
find that indeed they give the harmonic sums and that there is a one to one 
correspondence between them. As a consequence the investigation also leads 
to a rather simple algorithm for the inverse Mellin transform, even though 
in general the length of the resulting formulae requires a computer 
implementation for dealing with the great number of terms which are generated. 

All algorithms that we present have been programmed in the language of 
FORM~\cite{FORM}. The resulting procedures can be obtained from the second 
author.

\section{ Definitions. } 

The harmonic polylogarithms of weight $w$ and argument $x$ 
are identified by a set of $w$ indices, grouped into a 
$w$-dimensional vector $\vec{m}_w$ and are indicated by 
$\Hpl(\vec{m}_w;x)$. \\ 
More explicitly, for $w=1$ one defines 
\begin{eqnarray} 
  \Hpl(0;x) &=& \ln{x} \ ,          \nonumber\\ 
  \Hpl(1;x) &=& \int_0^x \frac{dx'}{1-x'} = - \ln(1-x) \ , \nonumber\\ 
  \Hpl(-1;x) &=& \int_0^x \frac{dx'}{1+x'} = \ln(1+x) \ . 
\labbel{eq:defineh1}
\end{eqnarray} 
For their derivatives, one has 
\begin{equation}
  \frac{d}{dx} \Hpl(a;x) = f(a;x) \ , 
\labbel{eq:derive1} 
\end{equation}
where the index $a$ can take the 3 values $0, +1, -1$ and the 
3 rational fractions $f(a;x)$ are given by 
\begin{eqnarray}
   f(0;x) &=& \frac{1}{x} \ , \nonumber\\
   f(1;x) &=& \frac{1}{1-x} \ , \nonumber\\
   f(-1;x) &=& \frac{1}{1+x} \ .
\labbel{eq:definef}
\end{eqnarray}
Note the (minor) asymmetry of \Eq{eq:defineh1}, in contrast with the higher 
symmetry of \Eq{eq:derive1}.

\noindent For $w > 1$, let us elaborate slightly the 
notation for the $w$-dimensional vectors $\vec{m}_w$. Quite in 
general, let us write
\begin{equation}
 \vec{m}_w = ( a, \vec{m}_{w-1} ) \ ,
\end{equation}
where $a=m_w$ is 
the leftmost index (taking of course one of the three values $0, 1, -1 
$), and $\vec{m}_{w-1}$ stands for the vector of the remaining  
$(w-1)$ components. Further, $\vec{0}_w$ will be the vector whose $w$ 
components are all equal to the index $0$. The harmonic 
polylogarithms of weight $w$ are then defined as follows:
\begin{equation}
\Hpl(\vec{0}_w;x) = \frac{1}{w!} \ln^w{x} \ ,
\labbel{eq:defh0}
\end{equation}
while, if $\vec{m}_w \neq \vec{0}_w$
\begin{equation}
\Hpl(\vec{m}_w;x) = \int_0^x dx' \ f(a;x') \ \Hpl(\vec{m}_{w-1};x') \ .
\labbel{eq:defn0}
\end{equation}
Quite in general the derivatives can be written in the compact form 
\begin{equation}
\frac{d}{dx} \Hpl(\vec{m}_w;x) = 
        f(a;x) \Hpl(\vec{m}_{w-1};x) \ , 
\labbel{derive} \end{equation}
where, again, \( a=m_w \) is the leftmost component of \( \vec{m_w} \). 
\par 
In analogy with \Eq{eq:defh0}, if $\vec{1}_w, \vec{(-1)}_w$ are the 
vectors whose components are all equal to $1$ or $-1$, we have by applying 
recursively the definitions 
\begin{eqnarray} 
  \Hpl(\vec{1}_w;x) = \frac{1}{w!} ( - \ln{(1-x)} )^w \ , \nonumber\\ 
  \Hpl(\vec{(-1)}_w;x) = \frac{1}{w!} \ln^w{(1+x)} \ . 
\labbel{eq:defh1-1}
\end{eqnarray} 
\par 
Let us now have a look at the first few values of the indices. 
For $w = 2$ one has the 9 functions
\begin{eqnarray}
  \Hpl(0,0;x) &=& \frac{1}{2!} \ln^2{x} \ , \nonumber\\
  \Hpl(0,1;x) &=& \int_0^x \frac{dx'}{x'} \Hpl(1;x')
            = - \int_0^x \frac{dx'}{x'} \ln(1-x') \ , \nonumber\\
  \Hpl(0,-1;x) &=& \int_0^x \frac{dx'}{x'} \Hpl(-1;x')
            =   \int_0^x \frac{dx'}{x'} \ln(1+x') \ , \nonumber\\
  \Hpl(1,0;x) &=& \int_0^x \frac{dx'}{1-x'} \Hpl(0;x')
            = \int_0^x \frac{dx'}{1-x'} \ln{x'} \ , \nonumber\\
  \Hpl(1,1;x) &=& \int_0^x \frac{dx'}{1-x'} \Hpl(1;x')
            = - \int_0^x \frac{dx'}{1-x'} \ln(1-x') \ , \nonumber\\
  \Hpl(1,-1;x) &=& \int_0^x \frac{dx'}{1-x'} \Hpl(-1;x')
            = \int_0^x \frac{dx'}{1-x'} \ln(1+x') \ , \nonumber\\
  \Hpl(-1,0;x) &=& \int_0^x \frac{dx'}{1+x'} \Hpl(0;x')
            = \int_0^x \frac{dx'}{1+x'} \ln{x'} \ , \nonumber\\
  \Hpl(-1,1;x) &=& \int_0^x \frac{dx'}{1+x'} \Hpl(1;x')
            = - \int_0^x \frac{dx'}{1+x'} \ln(1-x') \ , \nonumber\\
  \Hpl(-1,-1;x) &=& \int_0^x \frac{dx'}{1+x'} \Hpl(-1;x')
            = \int_0^x \frac{dx}{1+x'} \ln(1+x') \ . 
\labbel{eq:weight2int}
\end{eqnarray}
Those 9 functions can all be expressed in terms of logarithmic and 
dilogarithmic functions; indeed, if 
\begin{equation}
   \Li_2(x) = - \int_0^x \frac{dx'}{x'}\ln(1-x') 
\labbel{Li2} \end{equation}
is the usual Euler's dilogarithm, one finds
\begin{eqnarray}
  \Hpl(0,1;x) &=& \Li_2(x) \ , \nonumber\\
  \Hpl(0,-1;x) &=& - \Li_2(-x) \ , \nonumber\\
  \Hpl(1,0;x) &=& - \ln{x} \ln(1-x) + \Li_2(x) \ , \nonumber\\
  \Hpl(1,1;x) &=& \frac{1}{2!} \ln^2(1-x) \ , \nonumber\\
  \Hpl(1,-1;x) &=& \Li_2\left(\frac{1-x}{2}\right) - \ln2 \ln(1-x)
        -\Li_2\left(\frac{1}{2}\right) \ , \nonumber\\
  \Hpl(-1,0;x) &=& \ln{x} \ln(1+x) + \Li_2(-x) \ , \nonumber\\
  \Hpl(-1,1;x) &=& \Li_2\left(\frac{1+x}{2}\right) - \ln2 \ln(1+x)
        -\Li_2\left(\frac{1}{2}\right) \ , \nonumber\\
  \Hpl(-1,-1;x) &=& \frac{1}{2!} \ln^2(1+x) \ .
\labbel{eq:weight2}
\end{eqnarray}
Something similar happens for harmonic polylogarithms and Nielsen's 
polylogarithms of weight 3; that is no longer true however from 
weight 4 on. To make an example, 
\begin{equation}
   \Hpl(-1,0,0,1;x) = \int_0^x \frac{dx'}{1+x'} \Li_3(x') 
\labbel{eq:weight4ex}
\end{equation}
cannot be expressed in terms of Nielsen's polylogarithms 
of the same weight, even allowing for slightly more general arguments 
({\it i.e.} when considering, besides $x$, also $-x$, $(1+x)/2, 
(1-x)/2$ etc.). In other words, the set of the $3^w$ harmonic 
polylogarithms of weight $w$ is in general a much wider set of 
functions than the set of the Nielsen's polylogarithms.
\par 
It follows from the definition that if $\vec{m}_w \neq \vec{0}_w$ the 
hpl's vanish at $x=0$: 
\begin{equation}
  \Hpl(\vec{m}_w;0) = 0, \hskip 2truecm \vec{m}_w \neq \vec{0}_w \ . 
\labbel{eq:defh00}
\end{equation}
Likewise, if the leftmost index $m_w$ is not equal to 1, 
($m_w \neq 1$), $\Hpl(\vec{m}_w;1)$ is finite; it is also finite 
when $\vec{m}_w = 1$, but all the remaining indices $\vec{m}_{w-1} $ are 
zero, ($\vec{m}_{w-1} = \vec{0}_{w-1}$). In the remaining cases, 
{\it i.e. } $m_w = 1$ and $\vec{m}_{w-1} \ne \vec{0}_{w-1}$, 
$\Hpl(\vec{m}_w;x)$ has a logarithmic behaviour at $x =1$: more 
exactly, if the $p$ leftmost indices are all equal to $1$, 
$\Hpl(\vec{m}_w;x)$ behaves for $x\to 1$ as a combination 
of powers of $\ln(1-x)$ ranging from the maximum value $p$ 
down to $0$ (the maximum power is decreased to \( p-1 \) if the 
remaining $w-p$ indices are all equal to zero; 
the study of the detailed logarithmic behaviours at $x=0,1$ will be carried 
out in Section 3). 
\par 
In dealing with specific cases and except for the smallest values of $w$, 
specifying explicitly all the components of $\vec{m}$ becomes quite 
cumbersome, so that a more compact notation is welcome. 
In the case that we ignore the functions of which the last index is zero we 
can use the same compactified notation as in ref~\cita{HS}. This is to say 
that, proceeding from right to left, all zeroes are simply eliminated by 
adding at the same time one to the absolute value of the previous index 
to the right, as in
\begin{equation}
\labbel{eq:notchange}
    \Hpl(0,0,1,0,-1;x) = \Hpl_{3,-2}(x) . 
\end{equation}
In terms of this notation and excluding, as already stated, the cases in 
which the rightmost index is zero, one can formulate the following:

\noindent {\sl theorem}: If $m_1 \ne 0$ one has
\begin{equation}
\labbel{eq:minsign}
 \Hpl_{m_p,\cdots,m_1}(-x) = \sign(p) \Hpl_{-m_p,\cdots,-m_1}(x)\ .
\end{equation}
The proof goes by induction and follows rather trivially from the 
definition of $\Hpl$. In the case that we use the notation in which the 
$m_i$ only have the value $0,1,-1$ the power of $-1$ is the number of indices 
that are not zero.
\\
In general we will write the indices of the $\Hpl$-functions as subscripts 
when we use the notation of the {\sl r.h.s.} of \Eq{eq:notchange}, while we 
will use the notation of the {\sl l.h.s.} when the indices are supposed to 
have the values $0,1,-1$ only.
In that last notation, to see the relation with the polylogarithms of 
Nielsen $S_{n,p}(x)$, defined in \cita{Nielsen}, let us indicate with 
$\vec{0}_n, \vec{1}_p$ as usual, two $n$-dimensional and 
$p$-dimensional vectors whose components are all equal to $0$ and $1$ 
respectively; one then has 
\begin{equation}
  S_{n,p}(x) = \Hpl( \vec{0}_n, \vec{1}_p ; x) \ . 
\labbel{Nielsen}
\end{equation}

As an obvious extension of the terminology, the product of two 
$\Hpl$-functions of weight $w_1$ and $w_2$ will be said to have total 
weight $w=w_1+w_2$. In the following we will often encounter homogeneous 
``identities of weight $w$", {\it i.e.} relations (or identities) involving 
the sum of several terms, where each term is equal to the product of an 
integer or rational fraction times a $\Hpl$-function of weight \( w \) or a 
product of several $\Hpl$-functions separately of lower weight but with 
total weight  \( w \). \par While the $\Hpl$-functions of weight $w$ are 
linear independent, the same is not true for the wider set of all the 
homogeneous expressions of weight $w$. The redundance can be used for 
establishing a number of (homogeneous) identities expressing a 
$\Hpl$-function of some argument and weight $w$ as a homogeneous expression 
of the same weight involving $\Hpl$-functions of the same or of related 
arguments (including constant arguments, such as for instance \( +1 \) or 
\( -1 \)). The identities can be useful, typically, for exhibiting 
explicitly the behaviour at particular points (such as the logarithmic 
behaviour at $0$ or $\pm 1$) or for obtaining relations between 
$\Hpl$-functions of special arguments. Quite in general, while establishing 
such identities can be more or less wearisome, there is almost always a 
straightforward ``standard method" for checking a given identity: one first 
verifies that the identity holds for a particularly convenient choice of 
the variable (or variables) and then differentiate it with respect to one 
of the arguments. In so doing one obtains another relation, however of 
lower weight, according to \Eq{derive}; the procedure can be iterated until 
a relation of weight $1$ is eventually obtained, whose check is trivial 
(because the $\Hpl$-functions of weight $1$ are just logarithms). \par 
Likewise, also the mathematical constants corresponding to the particular 
values of the $\Hpl$-function of weight $w$ (such as the values at $x=1$ 
when finite) can be given the same weight $w$. Those values, at $x=\pm1$ or 
other simple arguments, are of particular interest by themselves, as it 
turns out that they can be expressed in terms of a very small number of 
mathematical constants, such as Riemann \( \zeta \)-functions, \( \ln2 \) 
etc. We will see that they are connected to the sums to infinity of 
ref~\cite{HS} which have been systematically evaluated and tabulated\footnote{
in ref~\cite{HS} this was done only to weight $=7$} by one of us (J.V.) to 
weight $=9$ and can be evaluated basically to any weight, given enough 
computer resources\footnote{An alternative method to obtain the finite 
constants consists of their numerical evaluation to high precision and then 
fitting them to a presumed basis. Using this method 
Broadhurst~\cite{DBprivate} has evaluated all finite objects at weight $=9$ 
and some objects at the weights $10$ and $11$}. In similar ways 
these sums have been evaluated under the name of Euler/Zagier sums by the 
authors of ref~\cite{BBB}. Hence, whenever $\Hpl$-functions at $x=1$ will 
appear in this paper they can be regarded as known from ref~\cite{HS} or 
ref~\cite{BBB}, provided their weight is not too large. It will also be 
shown that one may alternatively consider them as unknown constants, to be 
expressed in terms of that much smaller number of mathematical constants by 
systematically exploiting the many identities among $\Hpl$'s of various 
arguments established in the rest of this paper.
%

\section{ Identities between functions of the same argument. } 

Let us start by the integration by parts (ibp) identities. From the very 
definition,
\begin{eqnarray} 
 \Hpl(m_1\cdots m_q;x) &\!=\! & 
  \int_0^x dx'\ f(m_1;x')\Hpl(m_2\cdots m_q;x') 
  \nonumber \\ &\!=\!& 
  \Hpl(m_1;x)\Hpl(m_2\cdots m_q;x) 
 -\!\int_0^x dx'\ \Hpl(m_1;x')f(m_2;x')\Hpl(m_3\cdots m_q;x') 
 \nonumber \\ &\!=\!& 
  \Hpl(m_1;x)\Hpl(m_2\cdots m_q;x) 
  -\Hpl(m_2m_1;x)\Hpl(m_3\cdots m_q;x) 
  \nonumber \\ &\!+\!& 
  \Hpl(m_3m_2m_1;x)\Hpl(m_4\cdots m_q;x) 
 -\cdots -\sign(p)\Hpl(m_q\cdots m_1;x) \ . 
\labbel{eq:ibp} \end{eqnarray} 
The above identity can be immediately verified, independently of its 
derivation, by the `standard methods': it holds at $x = 0$; when 
differentiating with respect to $x$, one obtains a number of terms 
which are immediately seen to cancel out pairwise; therefore, the relation 
is true. 
This relation shows that in the case that $\vec{m}_q$ is symmetric and $q$ 
is even the $\Hpl$-function reduces to products of lower weight functions. 
In general the relation can be used when it is important to reduce the 
number of $\Hpl$-functions with the highest weight as much as possible.
\par 
Another important set of identities expresses the product of any two 
$\Hpl$-functions of weight \( w_1 \) and \( w_2 \) as a linear combination 
of $\Hpl$-functions of weight \( w=w_1+w_2 \). Let us start from the 
case \( w_1 = 1 \); the identity reads 
\begin{eqnarray}
   \Hpl(a;x) \Hpl(m_p,\cdots,m_1;x) &=& \Hpl(a,m_p\cdots,m_1;x) \nonumber\\ 
      &+& \Hpl(m_p,a,m_{p-1}\cdots,m_1;x) \nonumber \\ 
      &+& \Hpl(m_p,m_{p-1},a,m_{p-2}\cdots m_1;x) \nonumber \\ 
      &+& \cdots \nonumber \\ 
      &+& \Hpl(m_p,\cdots,m_1,a;x) \ . 
\labbel{eq:single} \end{eqnarray}
It can be established by induction in $p$. For $p =1$ it is almost 
trivial, corresponding to \Eq{eq:ibp} for $q=2$. Assume then that 
it holds for $p-1$; take the identity for $p-1$, multiply 
by $f(m_p;x)$ and integrate over $x$. In the {\sl r.h.s.} we 
can do the integral and obtain all necessary terms except for the one 
starting with $a$. The {\sl \l.h.s.} can be integrated by parts 
to give the proper {\sl l.h.s.} term plus 
another term that can be integrated and gives indeed the missing term. 
This completes the proof.
\par 
Again, once established the identity can also be 
verified by the `standard method': it holds at $x=0$; 
the $x$-derivative consists of two groups of terms, a first 
group with the coefficient $f(a;x)$ contains just two terms which 
cancel out immediately, plus a second group proportional to $f(m_p;x)$, 
which is nothing but the same relation at level $p-1$, so that the 
procedure can be repeated $p$ times until everything cancels out.
\par
There is only one complication with \Eq{eq:single}. This 
concerns points in which one of the objects involved is divergent. Hence 
one cannot apply this equation for $x=1$ in the case that either $a=1$ or 
$m_p=1$. This is explained 
better in the section on the algebraic properties.
\par
\Eq{eq:single} can be generalized to the product of two 
$\Hpl$-functions $\Hpl(\vec{p};x) \Hpl(\vec{q};x)$; if $p,q$ are 
the dimensions of $\vec{p},\vec{q}$ (or, which is the same, the weights 
of the two $\Hpl$-functions), the product is equal to the sum of $ 
(p+q)!/p!q!$ terms, each term being an $\Hpl$-function of weight $(p+q)$ 
with coefficient $+1$, obtained by choosing $p$ indices in all possible 
ways (hence the binomial coefficients) and filling them from left to 
right with the components of \( \vec{p} \) without changing their order, 
while the remaining $q$ places contain the components of $\vec{q}$, 
again without altering their order. This can be expressed with the formula 
\begin{eqnarray}
 \Hpl(\vec{p};x)\Hpl(\vec{q};x) & = &
  \sum_{\vec{r} = \vec{p}\uplus \vec{q}} \Hpl(\vec{r};x)
\labbel{eq:halgebra} \end{eqnarray}
in which $\vec{p}\uplus \vec{q}$ represents all mergers of $\vec{p}$ and 
$\vec{q}$ in which the relative orders of the elements of $\vec{p}$ and 
$\vec{q}$ are preserved.
\par 
As an example, for $p=2, \vec{p}=(a,b)$ and $q=3, \vec{q}=(r,s,t)$ one has
\begin{eqnarray} 
   \Hpl(a,b;x) \Hpl(r,s,t;x) &=& \Hpl(a,b,r,s,t;x) 
                        + \Hpl(a,r,b,s,t;x) \nonumber\\ 
                       &+& \Hpl(a,r,s,b,t;x) + \Hpl(a,r,s,t,b;x) \nonumber\\ 
                       &+& \Hpl(r,a,b,s,t;x) + \Hpl(r,a,s,b,t;x) \nonumber\\ 
                       &+& \Hpl(r,s,a,b,t;x) + \Hpl(r,a,s,t,b;x) \nonumber\\ 
                       &+& \Hpl(r,s,a,t,b;x) + \Hpl(r,s,t,a,b;x) \ , 
\labbel{eq:prod2-3} \end{eqnarray}
as can be easily checked, again, by the `standard method'.
\par
The product identities \Eq{eq:halgebra} can be used to single out the terms in 
$\ln(x)$ from $\Hpl$-functions whose indices have trailing (or rightmost) 
indices equal to zero (as we will see in the next 
section $\Hpl$-functions with no trailing zeroes can be expanded in series 
of $x$ around $x=0$, while $\Hpl$-functions with trailing zeroes develop 
logarithmic singularities at that point). For $a=0$ in \Eq{eq:single}, 
recalling \( \Hpl(0;x) = \ln(x) \), \Eq{eq:defh0} and 
\Eq{eq:defineh1} one obtains
\begin{eqnarray}
 \Hpl(m_1,\cdots,m_p,0;x) & = &
   \ln(x) \Hpl(m_1,\cdots,m_p;x)
   -\Hpl(0,m_1,\cdots,m_p;x)
  \nonumber \\ &&
   -\Hpl(m_1,0,m_2,\cdots,m_p;x)
   -\cdots -\Hpl(m_1,\cdots,m_{p-1},0,m_p;x)\ .
\labbel{eq:lnx} \end{eqnarray}
In the case that $m_p$ is also zero we can move the last term to the left, 
divide by two and then use again \Eq{eq:single} for all the 
other terms, thus obtaining an identity which extracts the logarithmic 
singularities due to $2$ trailing zeroes. By suitably repeating 
the procedure as many times as needed, we can extract in general all the 
powers of $\ln(x)$ from the generic $\Hpl$-function. A couple of examples, 
if $a,b$ are any non-zero indices, are 
\begin{eqnarray}
   \Hpl(a,b,0,0;x) &=& \Hpl(0,0;x) \Hpl(a,b;x) \nonumber\\ 
                &-& \Hpl(0;x)\biggl( \Hpl(a,0,b;x) + \Hpl(0,a,b;x) \biggr) \nonumber\\ 
   &+& \Hpl(a,0,0,b;x) + \Hpl(0,a,0,b;x) + \Hpl(0,0,a,b;x) \ , \nonumber\\ 
   \Hpl(a,b,0,0,0;x) &=& \Hpl(0,0,0;x) \Hpl(a,b;x) \nonumber\\ 
              &-& \Hpl(0,0;x)\biggl( \Hpl(a,0,b;x) + \Hpl(0,a,b;x) \biggr) \nonumber\\ 
   &+& \Hpl(0;x) \biggl( \Hpl(a,0,0,b;x) + \Hpl(0,a,0,b;x) + \Hpl(0,0,a,b;x) 
                                           \biggr) \nonumber\\ 
   &-& \biggl( \Hpl(a,0,0,0,b;x) + \Hpl(0,a,0,0,b;x) \nonumber \\ &&
    + \Hpl(0,0,a,0,b;x) + \Hpl(0,0,0,a,b;x) \biggr)
\end{eqnarray}
\par In the same way one can use the product identities, \Eq{eq:halgebra} 
for extracting the terms singular as powers of \( \ln(1-x) \), or 
equivalently of $H(1;x)$ according to \Eq{eq:defineh1}, 
around \( x=1 \) from the $\Hpl$-functions whose leading 
(or leftmost) indices are equal to 
1. If \( a=1 \) \Eq{eq:single} can be rewritten as
\begin{eqnarray}
 \Hpl(1,m_1,\cdots,m_p;x) & = &
  \Hpl(1;x) \Hpl(m_1,\cdots,m_p;x)
  -\Hpl(m_1,1,m_2\cdots,m_p;x)
  \nonumber \\ & - &
  \Hpl(m_1,m_2,1,\cdots m_p;x)
 -\cdots -\Hpl(m_1,\cdots,m_{p-1}m_p,1;x)\ .
\labbel{eq:ln(1-x)} \end{eqnarray}
If \( m_1 \) has also the value \( 1 \) we can take the second term of the 
{\it r.h.s.} to the left, divide by two and obtain an identity to be used 
when the first \( 2 \) indices are both equal to \( 1 \) and so on. Let 
us show a couple of examples in the case of two indices \( a,b \) not 
equal to \( 1 \): 
\begin{eqnarray} 
   \Hpl(1,1,a,b;x) &=& \Hpl(1,1;x) \Hpl(a,b;x) \nonumber\\ 
                &-& \Hpl(1;x)\biggl( \Hpl(a,1,b;x) + \Hpl(a,b,1;x) \biggr) \nonumber\\ 
   &+& \Hpl(a,1,1,b;x) + \Hpl(a,1,b,1;x) + \Hpl(a,b,1,1;x) \ , \nonumber\\ 
   \Hpl(1,1,1,a,b;x) &=& \Hpl(1,1,1;x) \Hpl(a,b;x) \nonumber\\ 
              &-& \Hpl(1,1;x)\biggl( \Hpl(a,1,b;x) + \Hpl(a,b,1;x) \biggr) \nonumber\\ 
   &+& \Hpl(1;x) \biggl( \Hpl(a,1,1,b;x) + \Hpl(a,1,b,1;x) + \Hpl(a,b,1,1;x) 
                                           \biggr) \nonumber\\ 
   &-& \biggl( \Hpl(a,1,1,1,b;x) + \Hpl(a,1,1,b,1;x) \nonumber \\ &&
  + \Hpl(a,1,b,1,1;x) 
                              + \Hpl(a,b,1,1,1;x) \biggr) \ ;
\end{eqnarray}
the structure is very much the same as in the equations 
for extracting the \( \ln(x) \) singularities related to the trailing 
zeroes.

It is to be noted that the two procedures -- the ``extraction" of leading 
\( 1 \)'s and trailing \( 0 \)'s -- can be combined, to give, for instance 
\begin{eqnarray} 
 \Hpl(1,1,-1,0;x) &=& \frac{1}{2}\Hpl(-1;x)\Hpl(0;x)\Hpl^2(1;x) 
                -  \Hpl(-1,1;x)\Hpl(0;x)\Hpl(1;x) \nonumber \\ 
    &+&  \Hpl(-1,1,1;x)\Hpl(0;x) - \frac{1}{2}\Hpl(0,-1;x)\Hpl(1;x)\Hpl(1;x) \nonumber\\
    &+&  \Hpl(0,-1,1;x)\Hpl(1;x) -  \Hpl(0,-1,1,1;x) \ , \nonumber\\ 
 \Hpl(1,1,0,0,0;x) &=& \frac{1}{12} \Hpl^3(0;x) \Hpl^2(1;x) \nonumber\\ 
    &-&  \Hpl(0,0,0,1;x)\Hpl(1;x) +  \Hpl(0,0,0,1,1;x) \nonumber\\ 
    &+&  \Hpl(0,0,1;x)\Hpl(0;x)\Hpl(1;x) -  \Hpl(0,0,1,1;x)\Hpl(0;x) \nonumber\\ 
    &-&  \frac{1}{2} \Hpl(0,1;x)\Hpl^2(0;x)\Hpl(1;x)     \nonumber\\ 
    &+&  \frac{1}{2} \Hpl(0,1,1;x)\Hpl^2(0;x) \ . 
\labbel{ex11-10}
\end{eqnarray}
Therefore, one can always express a $\Hpl$-function with leading \( 1 \)'s 
and trailing \( 0 \)'s in terms of products of powers of \( H(0;x) \) and 
\( H(1;x) \), which exhibit the logarithmic singularities in those points, 
and of other ``irreducible" $\Hpl$'s, {\it i.e.} $\Hpl$'s whose first index 
is not \( 1 \) and the last index is not \( 0 \) and therefore is finite at 
both \( x=1 \) and \( x=0 \).

We can push further this kind of reduction, 
by writing all the possible product identities \Eq{eq:halgebra} and the 
integration by part identities \Eq{eq:ibp} and using them for expressing 
as many as possible $\Hpl$'s of weight \( w \) and ``unwanted" indices in 
terms of products of a ``minimal" set of $\Hpl$'s of lower weight and 
``accepted" indices. It is to be noted that the number of the $\Hpl$'s in the 
``minimal" set is fixed, but their choice is not unique, even if the condition 
of the extraction of the leading \( 1 \)'s and trailing \( 0 \)'s is 
imposed. It is easily seen that at weight \( w \) the number of 
relations is nothing but the total number of the different products of 
$\Hpl$'s of lower weight and with total weight \( w \). These relations are 
independent when all $\Hpl$-functions of lower weight belong to their 
respective ``minimal sets". 
It is to be observed, in any case, that the above ``reduction" involves 
only different rearrangements, without any modification, of the set of 
indices which appear in the original $\Hpl$, 
\par
An explicit calculation gives the set sizes of table~\ref{tab:basis}.
\begin{table}[htb]
\centering
\begin{tabular}{r|rrr}
Weight & Full basis & Irreducible set & Minimal set \\ \hline
  2    &        9   &           4     &        3    \\
  3    &       27   &          12     &        8    \\
  4    &       81   &          36     &       18    \\
  5    &      243   &         108     &       48    \\
  6    &      729   &         324     &      116    \\
  7    &     2187   &         972     &      312    \\
  8    &     6561   &        2916     &      810
\end{tabular}
\caption{\label{tab:basis}\sl Sizes of the various bases}
\end{table}

The use of the full basis in which each term has only a single 
$\Hpl$-function gives a unique expression in a rather simple way. This is 
also the preferred representation when higher weights have to be built up 
by successive integration. Expressions can also be given in terms of the 
irreducible set in a relatively easy way. This form is preferred when one 
has to avoid problems with divergencies. It can also be convenient when 
establishing identities for related arguments. The use of the minimal set is 
particularly convenient for the numerical evaluation of the 
$\Hpl$-functions, when a large number of them has to be evaluated in the 
same point. It should also be noted that the use of a minimal set is 
relatively easy for the lower weights (at weight 3 it requires only 4 
substitutions) while for higher weights it will much less straightforward.

%
\section{Power series expansions}

In general the function $\Hpl_{\vec{m}}(x)$ does not have a regular Taylor 
series expansion. This is due to the effect that trailing zeroes in the 
index field may cause powers of $\ln(x)$. Hence the proper expansion is one 
in terms of both $x$ and $\ln(x)$. Let us first have a look at what happens 
when there are no logarithms. We will use now the other notation for the 
indices. In that case we have:
\begin{eqnarray}
 \Hpl_1(x) & = & \sum_{i=1}^\infty \frac{x^i}{i}
  \nonumber \\
 \Hpl_{-1}(x) & = & -\sum_{i=1}^\infty \frac{\sign(i)x^i}{i}
\end{eqnarray}
and assuming\footnote{Because of the linearity of the problem the 
presence of more than one term, each with a different $S_{\vec{n}}$ would 
not make much of a difference 
in the following considerations.} that
\begin{eqnarray}
 \Hpl_{\vec{m}}(x) & = & \sum_{i=1}^\infty \frac{\sigma^i x^i}{i^a}
   S_{\vec{n}}(i)
\end{eqnarray}
in which $\sigma = \pm 1$ one can write the relations
\begin{eqnarray}
\label{eq:recsum}
 \Hpl_{0,\vec{m}}(x) & = & \sum_{i=1}^\infty \frac{\sigma^i 
   x^i}{i^{a+1}} S_{\vec{n}}(i)
  \nonumber \\
 \Hpl_{1,\vec{m}}(x) & = &
   \sum_{i=1}^\infty \frac{x^i}{i} S_{\sigma a,\vec{n}}(i\minus 1)
  \nonumber \\ & = &
   \sum_{i=1}^\infty \frac{x^i}{i} S_{\sigma a,\vec{n}}(i)
   -\sum_{i=1}^\infty \frac{\sigma^i x^i}{i^{a+1}} S_{\vec{n}}(i)
  \nonumber \\
 \Hpl_{-1,\vec{m}}(x) & = &
   -\sum_{i=1}^\infty \frac{\sign(i)x^i}{i} S_{-\sigma a,\vec{n}}(i\minus 1)
  \nonumber \\ & = &
   -\sum_{i=1}^\infty \frac{\sign(i)x^i}{i} S_{-\sigma a,\vec{n}}(i)
   +\sum_{i=1}^\infty \frac{\sigma^i x^i}{i^{a+1}} S_{\vec{n}}(i)
\end{eqnarray}
At this point one could argue what is the better definition of the nested 
sums. A definition of the type
\begin{eqnarray}
 Z_{a,\vec{m}}(n) & = & \sum_{i=1}^n \frac{Z_{\vec{m}}(i\minus 1)}{i^a}
\end{eqnarray}
will give only a single term in the expansion and is favored in the 
mathematical literature, because there one is mainly concerned with sums to 
infinity. For finite values of $n$ however this definition has the 
unelegant aspect that when $\vec{m}$ has $k$ components that are not zero, 
the value of $Z_{a,\vec{m}}(n)$ is zero for $n \le k$. We will mostly 
follow the conventions of ref~\cita{HS} in which we use the definition:
\begin{eqnarray}
 S_{a,\vec{m}}(n) & = & \sum_{i=1}^n \frac{S_{\vec{m}}(i)}{i^a}
\end{eqnarray}
In this notation one has the property $S_{\vec{m}_k}(1) = 
\prod_{i=1}^k \sigma_i$ with $\sigma_i$ being the sign of $m_i$. These two 
notations will be referred to as $Z$-notation and $S$-notation 
respectively. 
The conversion from one notation to the other is not really 
very complicated if one realizes that $\sum_{j=1}^{i-1} = \sum_{j=1}^i - 
\delta_{ij}$. Hence the `leading' term has the same index field and the 
correction terms have fewer indices in which some adjacent indices may have 
been combined. For $k$ nonzero indices there are in total $2^{k-1}-1$ 
correction terms.

The fact that trailing zeroes in the index field are responsible for powers 
of $\ln(x)$ can be seen easily now. Because
\begin{eqnarray}
 \frac{1}{k!}\int^xdx\ x^m\ln^k(x) & = &
  x^{m+1}\sum_{\kappa=0}^k \frac{\sign(k-\kappa)}{\kappa!}
   \frac{\ln^\kappa(x)}{(m+1)^{k-\kappa+1}}
\end{eqnarray}
we see that once we start with $\Hpl(\vec{0}_k;x)$, the subsequent 
integrations due to other indices (the first of them not being zero of 
course, and factors $1/(1\pm x)$ being expanded in $x$) that come to the 
left of the $\vec{0}_k$ will always leave terms with at most $k$ powers of 
$\ln(x)$ and there will be a term with $k$ of those powers. Hence the 
trailing zeroes are responsible for powers of $\ln(x)$. Of course the exact 
dependence of $\ln(x)$ can be derived much easier by applying \Eq{eq:lnx} 
repeatedly 
till all trailing zeroes have been removed. This gives an expansion in 
terms of powers of $\ln(x)$ and $\Hpl$-functions that are of the type we 
have just studied and hence can be expanded in $x$. It is however also 
possible to work one's way through the integrals and the various expansions. 
This is much more work and leads eventually to the same 
result. Hence we have omitted this derivation.

If we compare the $\Hpl$-function with the multidimensional polylogarithm 
in ref~\cite{BBBL} we may notice that this function can be 
rewritten into the following expansion:
\begin{eqnarray}
\lambda(^{z_1\cdots z_k}_{b_1\cdots b_k}) & = &
 \sum_{\nu_1 > \nu_2 >\cdots > \nu_k > 0}^\infty
 \prod_{j=1}^k \frac{b_{j-1}^{\nu_j}}{\nu_j^{z_j}b_j^{\nu_j}}
\end{eqnarray}
with $b_0 = 1$. These functions do not contain powers of $\ln(b_i)$ and 
hence they cannot represent all $\Hpl$-functions. If we restrict ourselves 
to $\Hpl$-functions without trailing zeroes one can write the terms in the 
expansion of these $\Hpl$-functions as
\begin{eqnarray}
 \sum_{\nu_1 > \nu_2 >\cdots > \nu_k > 0}^\infty
 x^{\nu_1}\prod_{j=1}^k \frac{\sigma_j^{\nu_j}}{\nu_j^{z_j}}
\end{eqnarray}
if we use $Z$-sums and 
\begin{eqnarray}
 \sum_{\nu_1 \ge \nu_2 \ge\cdots \ge \nu_k \ge 1}^\infty
 x^{\nu_1}\prod_{j=1}^k \frac{\sigma_j^{\nu_j}}{\nu_j^{s_j}}
\end{eqnarray}
if we use $S$-sums. Hence it is clear that the $\Hpl$-functions without 
trailing zeroes are special cases of the multidimensional polylogarithms 
with $b_i = \pm 1/x$. For the computation of Feynman diagrams we do however 
need the $\Hpl$-functions with trailing zeroes because of the presence of 
the logarithms (see for instance ref~\cita{Zijlstra}).

There is another interesting observation in the expansion. Considering that 
the expansion of an $\Hpl$-function with no trailing zeroes gives terms of 
the type
\begin{eqnarray}
 \sum_{x=1}^\infty x^i \frac{\sigma^i S_{\vec{m}}(i)}{i^a}
   \nonumber
\end{eqnarray}
one can introduce another sum by dividing by either $1\plus x$ or $1\minus 
x$ and obtain:
\begin{eqnarray}
 \sum_{x=1}^\infty x^i \frac{\sigma^i S_{\vec{m}}(i)}{i^a}
  & = & (1\minus x) \sum_{x=1}^\infty x^i S_{\sigma a,\vec{m}}(i)
  \nonumber \\
  & = & (1\plus x) \sum_{x=1}^\infty x^i \sign(i) S_{-\sigma a,\vec{m}}(i)
\end{eqnarray}
At times this notation is more convenient. One should however remember that 
this notation breaks down at either $x=1$ or at $x=-1$, depending on the 
particular form used.

Finally we notice that for $x=1$ we have that
\begin{eqnarray}
 \sum_{x=1}^\infty x^i \frac{\sigma^i S_{\vec{m}}(i)}{i^a}
  & \rightarrow & S_{\sigma a,\vec{m}}(\infty)
\end{eqnarray}
and hence the values of the $\Hpl$-functions in $x=1$ are related to the 
values of the $S$-sums in infinity. The trailing zeroes do not cause 
essential problems because when those functions are first written in terms 
of powers of $\ln(x)$ these logarithms vanish in $x=1$ and we keep only the 
terms with $\Hpl$-functions without trailing zeroes. For the numerical 
evaluation of these objects one can use the algorithms of ref~\cita{BBBL} 
that relate them effectively to combinations of $\Hpl$-functions in $x=1/2$ 
after the appropriate conversions. This is particularly interesting for the 
higher weights because up to weight 7, 8 or 9 it is still possible to obtain 
expressions in terms of a very small number of constants (see 
ref~\cita{BBB} and \cita{HS}), but beyond these weights this becomes too time 
consuming\footnote{Thus far the only known exact method to do this involves 
solving simultaneously for all $2\ 3^{w-1}$ $\Hpl$-functions in $x=1$. See 
also a previous footnote.} 
while an expansion in $x=1/2$ is sufficiently fast for nearly all 
numerical applications, provided that only a limited number of them is 
needed.
%

\section{ The algebra}

\label{sec:algebra}
The harmonic sums form an algebra~\cita{HS} in which the product of two sums 
with the same argument and having weights $w_1$ and $w_2$ respectively can 
be written as a sum of terms, each with a single sum of weight $w_1+w_2$. 
There are two sets of algebraic relations: the relations based on the 
shuffle algebra which hold for all values of the argument, and the 
relations based on the triangle theorem of ref~\cita{HS} which hold only 
for values in infinity, provided that not both harmonic sums are divergent. 
For the $\Hpl$-functions we have the general product formula based on 
\Eq{eq:halgebra}. This formula is related to the algebra of the harmonic 
sums, because the harmonic polylogarithms can be expressed in terms of series 
expansions in which the coefficients are harmonic sums: 
assume for the moment that neither $\vec{m}$ nor $\vec{n}$ have 
trailing zeroes. In that case we derive:
\begin{eqnarray}
    \Hpl_{a,\vec{m}_p}(x)\Hpl_{\vec{n}_q}(x) & = &
        \frac{1}{1-x} \sum_{i=1}^\infty \frac{S_{\vec{m}_p}(i)x^i}{i^a}
                \sum_{j=1}^\infty S_{\vec{n}_q}(j)x^j
\end{eqnarray}
in which one of the two powers of $1/(1-x)$ has been absorbed in the sum 
over $i$. By combining the powers of $x$ this formula can be rewritten as
\begin{eqnarray}
    \Hpl_{a,\vec{m}_p}(x)\Hpl_{\vec{n}_q}(x) & = &
        \frac{1}{1-x}
    \sum_{i=1}^\infty x^i \sum_{j=1}^i 
           \frac{S_{\vec{m}}(j)S_{\vec{n}}(i-j)}{j^a}\ .
\end{eqnarray}
Note that the inner sum can be done and gives a set of terms that are all 
single $S$ functions, even though the expression may not be very compact. 
It is called a triangle sum and an algorithm for it is given in one of the 
appendices of ref~\cita{HS}. It is also available as a procedure in the 
language of FORM~\cite{FORM}. As a result one obtains an expression which 
can be resummed and gives terms with single $\Hpl$-functions.

For $\Hpl$-functions in $x=1$ we have seen that they can be directly 
expressed in terms of harmonic sums in infinity. Therefore the general 
algebraic rules for those sums that are based on the shuffle algebra for 
harmonic sums can be 
applied. Hence we see a duality here: the general rules for the 
$\Hpl$-functions correspond to the special triangle rules for the harmonic 
sums, and the special rules for the $\Hpl$-functions in $x=1$ correspond to 
the general shuffle rules for the harmonic sums.

There is one complicating factor when values in $x=1$ are considered. Let 
us start with assuming that the basic divergence $\Hpl(1;1)$ can be used as 
a symbol. In the case of a `proper' limit procedure such things can be 
done and after the divergences cancel the finite result should be correct. 
This is called regularization. 
The general algebraic relations are based on the triangle sums, 
rather than on the shuffle algebra, and the triangle sums are not 
correct when both objects are divergent. The subleading terms will be 
incorrect. This can be illustrated easily:
\begin{eqnarray}
 \Hpl_1(x) & = & \sum_{i=1}^\infty \frac{x^i}{i}
  \nonumber \\
 (\Hpl_1(x))^2 & = & 2 \Hpl_{1,1}(x)
  \nonumber \\ & = &
    2 \sum_{i=1}^\infty 
      x^i(\frac{S_1(i)}{i}-\frac{1}{i^2})
  \nonumber \\
 \Hpl_1(1) & = & \lim_{x\rightarrow 1} \sum_{i=1}^\infty \frac{x^i}{i}
  \nonumber \\ & = & S_1(\infty)
  \nonumber \\
 \Hpl_{1,1}(1) & = & \lim_{x\rightarrow 1}
    2 \sum_{i=1}^\infty 
      x^i(\frac{S_1(i)}{i}-\frac{1}{i^2})
  \nonumber \\ & = & 2 S_{1,1}(\infty) - 2 S_2(\infty))
  \nonumber \\ & = & (S_1(\infty))^2 - S_2(\infty))
\end{eqnarray}
and we see that
\begin{eqnarray}
 (\lim_{x\rightarrow 1} \Hpl_1(x))^2 & \ne &
 \lim_{x\rightarrow 1} (\Hpl_1(x))^2\ .
\end{eqnarray}
The solution to this problem is to be found in $S$-space. There it is 
possible to regularize the infinite sums in a consistent way by replacing 
the sum to infinity by a sum to $M$ with $M$ very large but finite, then 
one can have the divergences cancel and finally take the limit $M\rightarrow 
\infty$. This does not correspond to anything one can do in $x$-space. 
Because the triangle theorem does not hold for two $S$-sums that are 
divergent, one cannot apply the regular algebraic relation for 
$\Hpl$-functions that are both divergent in $x=1$. 
Hence the proper algebraic relations at $x=1$ have to be derived by means 
of the shuffle algebra which holds for all $S$-sums:
\begin{eqnarray}
 (\lim_{x\rightarrow 1} \Hpl_1(x))^2 & = &
  (S_1(\infty))^2
  \nonumber \\ &=&
   2 S_{1,1}(\infty) - S_2(\infty)
  \nonumber \\ &=&
   2\lim_{x\rightarrow 1} \Hpl_{1,1}(x) + \lim_{x\rightarrow 1}\Hpl_2(x)
\end{eqnarray}
This way is consistent and will allow us to define the Mellin transform 
properly in one of the next sections. It involves the use of values in 
$x=1$.

Because of the use of different algebraic relations for $x\neq 1$ and 
$x=1$, it may happen that expressions look rather complicated, but the 
various algebraic relations between $\Hpl$-functions in $x=1$ could 
simplify the expressions considerably. However at the moment there is no 
known systematic method to apply these relations in such a way that one 
does not have to solve for all values in $x=1$ first. This way all such 
objects can be expressed in a minimal independent set of objects. 
Unfortunately there are very many of these objects for a given weight $w$ 
($2\ 3^{w-1}$) and even more relations and hence it is a formidable task to 
determine all values at $x=1$ in terms of a minimal set of constants when 
the weight is large. If the final answer is supposed to be finite one can 
however extract the powers of the basic divergences (they correspond to 
leading indices that are $1$) and hence still obtain a finite answer that 
can be evaluated numerically. The coefficients of the divergences can be 
checked to be zero numerically as well.

\section{ Identities between $\Hpl$-functions of related arguments.} 

In this section we will look at the identities which can be established 
for suitable changes of the argument. The common feature is that 
any $\Hpl$-function of weight $w$ and argument $x$ can be expressed as an 
homogeneous expression of the same weight $w$, involving either 
$\Hpl$-functions depending on a same argument, say \( t\), related to 
\( x \) by the considered change, or constants corresponding to 
$\Hpl$-functions of special constant values of the arguments 
(typically \( 1 \)). 
\par 
The simplest change of the argument is the change 
$x\rightarrow -x$. We have seen its effect already in \Eq{eq:minsign}.
\par 
Next is the relation between 
$\Hpl$-functions of $x^2$ and of $x$. Because $1+x^2$ is not a particularly 
interesting object we will have to exclude indices equal to -1 in the 
$\Hpl$-functions of $x^2$. Restricting the indices to only 
1 and 0, we can proceed recursively on the weight. For weight 1 we 
have from \Eq{eq:defineh1}:
\begin{eqnarray} 
      \Hpl(0;x^2) &=& 2\Hpl(0;x) \nonumber \\ 
      \Hpl(1;x^2) &=& \Hpl(1;x) - \Hpl(-1;x) \ , 
\labbel{eq:rel1}\end{eqnarray} 
so that the $\Hpl$'s of argument \( x^2 \) are expressed in terms of $\Hpl$'s 
of argument \( x \), as required. 
\par 
For $w > 1$, if $\vec{m}_w = \vec{0}_w$,
\begin{equation}
 \Hpl(\vec{0}_w;x^2) = 2^w \Hpl(\vec{0}_w;x) \ ;
\end{equation}
otherwise, if $\vec{m}_w = (a,\vec{m}_{w-1})$ for the two cases $a = 0$ 
and $a = 1$ we have, by using the change of variable $x' = t'^2$ 
\begin{eqnarray} 
\Hpl(0,\vec{m}_{w-1};x^2) & = &
  \int_0^{x^2} \frac{dx'}{x'} \Hpl(\vec{m}_{w-1};x')
  \nonumber \\ & = &
        2 \int_0^x \frac{dt'}{t'} \Hpl(\vec{m}_{w-1};t'^2) \\
\Hpl(1,\vec{m}_{w-1};x^2) & = &
   \int_0^{x^2} \frac{dx'}{1-x'} \Hpl(\vec{m}_{w-1};x') 
  \nonumber \\ & = &
        \int_0^x dt' \left( \frac{1}{1-t'} - \frac{1}{1+t'} \right) 
                                       \Hpl(\vec{m}_{w-1};t'^2) \ . 
\end{eqnarray} 
The expression of the \( \Hpl(\vec{m}_{w-1};t'^2) \) in terms of 
$\Hpl$'s of the same weight and argument \( t' \) is supposedly known 
(as we proceed recursively on the weight \( w \)); by substituting 
such expression and then using the very definition \Eq{eq:defn0} all the 
required \( x^2 \to x \) identities are obtained. 
An example of weight $w = 2$ is
\begin{eqnarray}
 \Hpl(0,1;x^2) & = &
        2 \int_0^x \frac{dt'}{t'} \Hpl(1;t'^2) \nonumber \\
  & = & 2 \int_0^x \frac{dt'}{t'}
    \left( \Hpl(1;t')-\Hpl(-1;t') \right) \nonumber \\
  & = & 2 \Hpl(0,1;x) - 2 \Hpl(0,-1;x)
\end{eqnarray}
and \Eq{eq:minsign} leads to the well known relation 
$\Li_2(x^2) = 2\Li_2(x)+2\Li_2(-x)$. We can observe here that a limited 
set of \( x^2 \to x \) identities could be written only for the Nielsen's 
polylogarithms corresponding to the \( \Hpl_n(x) \) in the notation of 
\Eq{eq:notchange}, while for the hpl's the set is wider; as an example, 
one can derive for $w=3$:
\begin{eqnarray}
 \Hpl(1,0,1;x^2) & = & 2\biggl(\Hpl(1,0,1;x)\minus\Hpl(-1,0,1;x)
   \minus\Hpl(1,0,-1;x)\plus\Hpl(-1,0,-1;x)\biggr)
\labbel{x^2wider}\end{eqnarray}
\par The next transformation of the argument we consider is $x \rightarrow 
1-x$ which applies again to a smaller set of Nielsen's polylogarithms. Like 
the previous transformation it is of interest only when there are no 
negative indices ($1+x \rightarrow 2-x$ is not something we can work with). 
Proceeding recursively on \( w \), as before, for \( w = 1 \) we have
\begin{eqnarray}
\Hpl(0;1-x) &=& -\Hpl(1;x) \nonumber \\
\Hpl(1;1-x) &=& -\Hpl(0;x).
\labbel{eq:rel1a}
\end{eqnarray}
The extension to higher weights requires a minimum of care. $\Hpl(a,
\vec{m}_{w-1};1-x) $ of weight \( w > 1 \), with the first index \( a \) 
equal to \( 0 \) or to \( 1 \) is the generic function. As discussed in 
Section 3, if \( a=1 \) the function can be expressed in terms of a reduced 
set of functions, where the leading index \( 1 \) is carried only by \( 
\Hpl(1;1-x) \), for which \Eq{eq:rel1a} holds; therefore, only the case in 
which the first index \( a \) is 0 is to be considered. In that case, the 
change of variable \( x'=1-t' \) gives
\begin{eqnarray}
\Hpl(0,\vec{m}_{w-1};1-x) &=& \int_0^{1-x} \frac{dx'}{x'} 
	\Hpl(\vec{m}_{w-1};x') \nonumber\\
	&=& \int_0^1 \frac{dx'}{x'}\Hpl(\vec{m}_{w-1};x')
	 - \int_{1-x}^1 \frac{dx'}{x'} \Hpl(\vec{m}_{w-1};x') 
		\nonumber\\
	&=& \Hpl(0,\vec{m}_{w-1};1)
	-\int_0^x \frac{dt'}{1-t'} \Hpl(\vec{m}_{w-1};1-t')\ , 
\end{eqnarray} 
where the constant \( \Hpl(0,\vec{m}_{w-1};1) \) is finite (it can be 
observed here that if the first index is $1$ one runs into 
the problem that $\Hpl(1,\vec{m}_{w-1};1)$ could be divergent). 
In the general case \( \Hpl(\vec{m}_{w-1};1-t') \) will not be irreducible. 
We can express it in terms of the $\Hpl$'s of an irreducible set of 
weight \( w-1 \), use the supposedly known \( x = 1-t \) identities of 
weight \( w-1 \) and finally obtain the required weight \( w \) identity 
by using the definition \Eq{eq:defn0}. 
As an example we have at weight 4 
\begin{eqnarray} 
 \Hpl(0,0,1,1;1-x) & = &
		\Hpl(0,0,1,1;1)-\Hpl(1;x)\Hpl(0,1,1;1)+\Hpl(1,1,0,0;x)
			\nonumber \\
	 &=& \Hpl(0,0,1,1;1) - \Hpl(0,1,1;1)\Hpl(1;x)
			- \Hpl(0,0,1,1;x) \nonumber \\
	 & + &\frac{1}{4} \Hpl^2(0;x)\Hpl^2(1;x) 
		   - \Hpl(0,1;x)\Hpl(0;x)\Hpl(1;x) \nonumber \\
     & + & \Hpl(0,0,1;x)\Hpl(1;x) + \Hpl(0,1,1;x)\Hpl(0;x)\ .
\labbel{1mxw4}\end{eqnarray} 
\par 
A transformation which applies to all the Nielsen's polylogarithms, 
\Eq{Nielsen} is 
\begin{equation}
    x = 1/y \ ; \hskip 2 truecm  y = 1/x \ ; 
\labbel{eq:xinv} 
\end{equation}
it will be shown that it applies as well to all the $\Hpl$-functions. 
Before continuing, let us recall that the Nielsen's polylogarithms have 
a (logarithmic) branch point at \( x = 1 \), but are otherwise analytic 
for smaller values of \( x \), including all the negative real axis; 
for studying the transformation \Eq{eq:xinv} it can be therefore convenient 
to establish the identities for negative values of \( x \), and then 
continue analytically to positive values. The analytic properties of the 
$\Hpl$-functions are more complicated. First of all, if the rightmost 
index is equal to \( 0 \), they have a branch point at \( x=0 \); that is 
not a problem, as we have already seen that we can express any $\Hpl$-function 
in terms of the functions of a reduced set where the trailing index \( 0 \) is 
carried only by powers of \( H(0;x)=\ln{x} \), whose analytic properties 
are well known. If the rightmost index is not \( 0 \) and all the indices 
are in general equal to \( 1 \) or \( 0 \), the $\Hpl$-functions have 
the same analytic properties as the Nielsen's polylogarithms; but if some 
of indices are equal to \( -1 \), a branch cut at \( x=-1 \) appears. 
Therefore, in the general case when indices equal to \( -1 \) are 
also present (and that is the case even of the reduced and minimal sets, see 
Section 3), there is no advantage in considering negative values of 
\( x \), so that we will start from the beginning with an argument 
equal to  \( x + \ieps \), where \( x \) is real and satisfies the 
constraints \( 0 \le x \le 1 \), while \( \epsilon \) is positive and 
infinitesimally small; correspondingly,
\begin{equation}
  y = 1/x - \ieps \ , 
\labbel{y}\end{equation}
{\it i.e.} the real part of \( y \) is also positive, but \( y \ge 1 \), 
while its infinitesimal imaginary part is negative. 
\par 
As in the previous cases, we will proceed by induction on the weight 
\( w \) of the $\Hpl$-functions. At \( w=1 \) we have 
\begin{eqnarray}
 \Hpl(0;y) &=& - \Hpl(0;x) \ , \nonumber \\ 
 \Hpl(1;y) &=& \Hpl(1;x) + \Hpl(0;x) - i\pi \ , \nonumber\\
 \Hpl(-1;y) &=& \Hpl(-1;x) - \Hpl(0;x) \ ; 
\labbel{eq:rel10}
\end{eqnarray}
the constant $\pi$ has appeared; it must be given weight $1$, so that 
all the formulas will remain homogeneous of the same weight. 
When continuing the above equations to negative values of \( x \), 
in the interval \( -1 \ge x \ge 0 \), \( H(0;x) = \ln(x+\ieps) \) will 
develop a positive imaginary part; in particular, one has
\begin{equation}
  \Hpl(0;-1) = i \pi \ , 
\labbel{-1+ieps}
\end{equation}
so that \(  \Hpl(1;-1) \) takes the real value \( - \ln2 \), as expected.
\par
For $w > 1$, $\vec{m}_w = (a,\vec{m}_{w-1}) $, we can proceed by 
induction, along the following lines
\begin{eqnarray}
 \Hpl(\vec{m}_w;y) &=& 
  \int_0^y dy'f(a;y') \Hpl(\vec{m}_{w-1};y') \nonumber\\
   &=& \int_0^1 dy'f(a;y') \Hpl(\vec{m}_{w-1};y')
   + \int_1^y dy'f(a;y') \Hpl(\vec{m}_{w-1};y') 
  \nonumber\\
   &=& \Hpl(\vec{m}_w;1) 
   + \int_x^1 \frac{dx'}{x'^2} f\left(a,\frac{1}{x'}\right) 
  \Hpl\left(\vec{m}_{w-1};\frac{1}{x'}\right) \ .
\labbel{eq:rel11} 
\end{eqnarray}
It is to be noted that one can assume that the first index \( a \) is 
different from \( 1 \); indeed, as seen in Section 3 any $\Hpl$-function 
of the form \( \Hpl(1,\vec{m}_{w-1};y) \) can be expressed in terms of a 
reduced set of functions, where the leading index \( 1 \) is carried only 
by powers of \( \Hpl(1;y) \), whose transformation is given by 
\Eq{eq:rel10}. For \( a \) different from \( 1 \), \( \Hpl(\vec{m}_w;1) \) 
is a finite constant and the above formulae are meaningful. 
One further finds 
\begin{eqnarray}
 \int \frac{dx'}{x'^2}f\left(0;\frac{1}{x'}\right) &=&
   + \int \ dx'\ \frac{1}{x'} \ , 
    \nonumber\\
 \int \frac{dx'}{x'^2} f\left(-1;\frac{1}{x'}\right) &=&
   + \int \ dx'\ \left(\frac{1}{x'} - \frac{1}{1+x'} \right) \ ;
    \nonumber 
\end{eqnarray} 
substituting in the {\it r.h.s.} of \Eq{eq:rel11} the identities (of 
weight \( w-1 \), and therefore known in an approach by induction) which 
express $\Hpl(\vec{m}_{w-1};y'=1/x')$ in terms of 
$\Hpl(\vec{m'}_{w-1};x')$, one obtains a combination of 
terms of the kind 
\[
  \int_x^1 \ dx'\ f(a;x') \Hpl_{\vec{m'}_{w-1}}(x') = 
   \Hpl(a,\vec{m'}_{w-1};1) - \Hpl(a,\vec{m'}_{w-1};x) \ , 
\] 
and the identities of weight \( w \) are established. As an example, we 
give the \( w=3 \) identity 
\begin{eqnarray} 
  \Hpl\left(0,-1,1; \frac{1}{x}-\ieps\right) &=& 
        - \Hpl(0,-1,1;x) + 2\Hpl(0,-1,1;1) \nonumber\\ 
    &&{\kern-50pt} + 2\Hpl(0,0,-1;x) - 2\Hpl(0,0,-1;1)
            + \Hpl(0,0,1;x)  - \Hpl(0,0,1;1) \nonumber\\ 
    &&{\kern-50pt} - \biggl( \Hpl(0,-1;x) + \Hpl(0,-1;1) + \Hpl(0,1;1) 
            \biggr) \Hpl(0;x) + \frac{1}{6}\Hpl^3(0;x)  \nonumber\\ 
    &&{\kern-50pt} - i\pi\left( \Hpl(-1;1)\Hpl(0;x) 
              + \frac{1}{2}\Hpl^2(0;x) - \Hpl(0,-1;x) 
              + H(0,-1;1) \right) \ . 
\labbel{ex:1/x} \end{eqnarray}

Another important set of identities, which is however valid for any 
set of indices and has no counterpart within the Nielsen's polylogarithms, 
applies to arguments $x$ and $t$ related by the transformation 
\begin{equation}
 x = \frac{1-t}{1+t}
  \labbel{eq:rel2} \ ,
\end{equation}
whose inverse is again 
\begin{equation}
 t = \frac{1-x}{1+x}
\labbel{eq:rel3} \ .
\end{equation}
Even in that case, it turns out that any $\Hpl$-function of weight 
$w$ and argument $x$ can be expressed as a homogeneous 
expression of weight $w$, involving $\Hpl$-functions of argument 
$t$, related to $x$ by \Eq{eq:rel2}, as well as constants 
corresponding to $\Hpl$-functions of argument $1$. 
The proof is, again, by induction on the weight. If $w = 1$, from 
the very definition \Eq{eq:defineh1} one immediately finds 
\begin{eqnarray} 
  \Hpl(0;x) &=& - \Hpl(1;t) - \Hpl(-1;t) \ , \nonumber\\ 
  \Hpl(1;x) &=& - \Hpl(0;t) - \Hpl(-1;1) + \Hpl(-1;t) \ , \nonumber\\ 
  \Hpl(-1;x) &=& \Hpl(-1;t) - \Hpl(-1;1) \ . 
\labbel{eq:rel4}
\end{eqnarray} 
For $w > 1$ and $\vec{m}_w = \vec{0}_w$ the result is 
trivially true, as can be verified by inspection; the same is true 
also for $\vec{m}_w = \vec{1}_w$ and $\vec{m}_w = \vec{-1}_w$. 
In the more general case, write $\vec{m}_w = (a,\vec{m}_{w-1})$; 
where the index $a$ takes the values $0,1,-1$. As discussed in Section 3, 
and already recalled for the \( x \to 1-x \) identities, if \( a=1 \) 
the function can be expressed in terms of a reduced set of functions, 
where the leading index \( 1 \) is carried only by \( \Hpl(1;x) \), 
for which \Eq{eq:rel4} holds. 
In the other two cases \( a=0,-1 \) the change of variable 
\begin{eqnarray}
 x' & = & \frac{1-t'}{1+t'} \nonumber
\end{eqnarray}
gives 
\begin{eqnarray}
 \Hpl(0,\vec{m}_{w-1};x) &=& \int_0^x \frac{dx'}{x'} 
   \Hpl(\vec{m}_{w-1};x') \nonumber \\ 
   &=& \Hpl(0,\vec{m}_{w-1};1)
   -\int_x^1 \frac{dx'}{x'} \Hpl(\vec{m}_{w-1};x') \nonumber \\
   &=& \Hpl(0,\vec{m}_{w-1};1) 
   -\int_0^t dt' \left( \frac{1}{1-t'} + \frac{1}{1+t'} \right) 
             \Hpl\left(\vec{m}_{w-1};\frac{1-t'}{1+t'} \right) \ ,
   \nonumber \\
 \Hpl(-1,\vec{m}_{w-1};x) &=& \int_0^x \frac{dx'}{1+x'} 
                              \Hpl(\vec{m}_{w-1};x') \nonumber\\ 
   &=& \Hpl(-1,\vec{m}_{w-1};1) -\int_0^t dt' \frac{1}{1+t'}  
             \Hpl\left(\vec{m}_{w-1};\frac{1-t'}{1+t'} \right) \ . 
\labbel{(1-t)/(1+t)} \end{eqnarray} 
At this point, one can substitute the relations already found to be valid 
at weight $w-1$, for expressing the functions 
\(  \Hpl\left(\vec{m}_{w-1};(1-t')/(1+t')\right) \) 
in terms of \( \Hpl \)'s of weight \( w-1 \) and argument \( t' \), 
and then perform the last integration in \( t' \) according to the 
definition \Eq{eq:defn0}. 
\par 
As an example, we give the following \( w=3 \) identity 
\begin{eqnarray} 
  \Hpl(-1,-1,1;x) &=& - \Hpl(0,-1,-1;t) + \Hpl(-1,-1,1;1) \nonumber\\ 
  &&{\kern-50pt} + \Hpl(0,-1;t)\Hpl(-1;t) 
        + \frac{1}{6}\Hpl^3(-1;t) 
        - \frac{1}{2}\Hpl^2(-1;t)\Hpl(0;t) \nonumber\\ 
  &&{\kern-50pt} - \frac{1}{2}\Hpl(-1;1)\Hpl^2(-1;t) 
        - \Hpl(-1,1;1)\Hpl(-1;t) \ . 
\labbel{tx3}\end{eqnarray} 

\section{ Identities between $\Hpl$'s and related functions.} 
Let us introduce a related set of functions \( \G(\vec{m}_w;x) \), where 
\( \vec{m}_w \)  has almost the same meaning as for the $\Hpl$'s, but the 
first index \( m_w \) is always equal to \( 1 \), {\it i.e.} 
\( \vec{m} = (1,\vec{m}_{w-1}) \), through the definitions 
\begin{equation}
  \G(1;x) = \intG 
\labbel{defG1}\end{equation}
for \( w=1 \) and 
\begin{equation}
  \G(1,\vec{m}_{w-1};x) = \intG \Hpl(\vec{m}_{w-1};t) 
\labbel{defGw}\end{equation}
for \( w>1 \). 
\par 
The \( \G(\vec{m}_w;x) \) are nothing but homogeneous combination of 
$\Hpl$-functions of weight \( w \). As by now usual, we will show it 
proceeding by induction on \( w \). For \( w=1 \), by performing explicitly 
the elementary integration we obtain from \Eq{defG1} 
\begin{equation}
  \G(1;x) = \Hpl(1;x) \ . 
\labbel{defGH1}\end{equation}
Next, assume that the identities are established for \( w \); put 
\( \vec{m} = (a,\vec{m}_{w-1}) \), and consider the functions of 
weight \( w+1 \) given by 
\begin{equation}
  \G(1,a,\vec{m}_{w-1};x) = \intG \Hpl(a,\vec{m}_{w-1};t) \ . 
\labbel{defGHw+1}\end{equation}
One can differentiate with respect to \( x \), then integrate by parts 
in \( t \), using of course \Eq{defGw} when relevant; considering for 
instance the case \( a=-1 \) one obtains 
\begin{eqnarray} 
  \frac{\partial}{\partial x}  \G(1,-1,\vec{m}_{w-1};x) &=& 
     \left[ f(-1,x) - f(0,x) \right] \G(1,\vec{m}_{w-1};x) \nonumber\\ 
    &+& \left[ f(-1,x) + f(1,x) \right] \Hpl(-1,\vec{m}_{w-1};1) \ . 
\labbel{derG-1}\end{eqnarray} 
Similarly, one has 
\begin{eqnarray} 
   \frac{\partial}{\partial x}  \G(1,0,\vec{m}_{w-1};x) &=& 
                    - f(0,x)  \G(1,\vec{m}_{w-1};x) \nonumber\\ 
    &+& f(-1,x) \Hpl(0,\vec{m}_{w-1};1)  \nonumber\\ 
   \frac{\partial}{\partial x}  \G(1,1,\vec{m}_{w-1};x) &=& 
     \left[ f(0,x) + f(1,x) \right] \G(1,\vec{m}_{w-1};x) \ . 
\labbel{derG01}\end{eqnarray} 
One can substitute the already obtained identities expressing 
\( \G(1,\vec{m}_{w-1};x) \) in terms of $\Hpl$'s of weight \( w \) and then 
integrate in \( x \) between \( 0 \) and \( x \) 
by using the very definition \Eq{eq:defn0} 
(according to \Eq{defGw} the \( \G \)-functions vanish at \( x=0 \)). 
The required identities of weight \( w+1 \) are then established. As an 
example, we give one of the identities of weight \( w=4 \) 
\begin{eqnarray} 
    \G(1,0,-1,1;x) &=& - \Hpl(0,-1,0,1;x) - \Hpl(0,-1,1,1;x) \nonumber\\ 
          &+& \Hpl(0,0,0,1;x) + \Hpl(0,0,1,1;x)              \nonumber\\ 
          &-& \Hpl(-1,1;1)\Hpl(0,-1;x) - \Hpl(-1,1;1)\Hpl(0,1;x) \nonumber\\ 
          &+& \Hpl(0,-1,1;1)\Hpl(1;x) \ .   
\labbel{Gw4ex}\end{eqnarray} 
\par 
In the same way one can work out the similar identities existing for 
several related classes of functions such as, for instance,  
\[   \intG \Hpl(\vec{a};t) \Hpl(\vec{b};xt)                 \] 
or 
\begin{equation}
           \int_0^1 dt f(a,t) \Hpl(\vec{a};t) \Hpl(\vec{b};xt) \ . 
\labbel{moreG}\end{equation}

\section{ Special values of the $\Hpl$'s and their numerical evaluation.} 

It is known that the Nielsen's polylogarithms for the special values of 
the arguments equal to \( +1, -1 \) and \( 1/2 \) can be expressed in terms 
of a few mathematical constants, typically Riemann \( \zeta \)-functions 
of integer arguments; the representations which they provide 
for those constants as definite integrals can be manipulated by means of 
integration by parts, changes of variables and the like providing the 
analytic values of a number of definite integrals of special interest. 
The same applies, and in much more systematic way, to the 
$\Hpl$-functions, thanks to the greater and wider sets of identities 
which they satisfy. 
\par 
In the case of the $\Hpl$'s, it is not necessary to consider as independent 
the values corresponding to the argument equal to \( -1 \); indeed, one can 
always express any 
$\Hpl$-function in terms of the reduced set of functions in which 
trailing indices equal to \( 0 \) are missing, so that by using 
\Eq{eq:minsign} one can replace a value at \( x=-1 \) with the value at 
\( x=1 \) of a related function. In analogy with the Nielsen's polylogarithms 
case, it is convenient to consider also the values at \( x=1/2 \) of 
the functions whose indices are equal to \( 0 \) or \( 1 \) ({\it i.e.} 
when the index \( -1 \) is missing). 
\par 
More specifically, one can consider: 
\begin{itemize}
\item the \( x^2 \to x \) identities, Eq.s(\ref{eq:rel1}-\ref{x^2wider}), 
for \( x=1 \) ; 
\item  the \( 1-x \to x \) identities, Eq.s(\ref{eq:rel1a}-\ref{1mxw4}). 
They can be used at \( x=1/2 \), providing a first set of identities for 
the values at \( x=1/2 \), but also at \( x=-1 \); in the second case, 
one gets values at \( x=2 \), which are converted into values at \( x=1/2 \) 
by using the \( x \to 1/x \) identities, Eq.s(\ref{eq:xinv}-\ref{ex:1/x}), 
as well as values at \( x=-1 \), which are converted into values at 
\( x=1 \) by \Eq{eq:minsign}; 
\item the just recalled \( x \to 1/x \) identities, 
Eq.s(\ref{eq:xinv}-\ref{ex:1/x}), at \( x=1 \) and \( x=-1 \), followed by 
the usual conversion  to \( x=1 \) through \Eq{eq:minsign}; 
\item the \( x \to (1-t)/(1+t) \) identities, Eq.s(\ref{eq:rel2}- 
\ref{tx3}), at \( x=0 \) corresponding to \( t=1 \) (they are automatically 
satisfied, by construction, at \( x=1, t=0 \)); 
\item one more set of identities is obtained by writing the identities 
between $\G$-functions and $\Hpl$-functions,discussed in Section 8, 
at the special value \( x=-1 \), by using the relation, which follows from 
the definition \Eq{defGw} 
\[ \G(1,\vec{m};-1) = \Hpl(-1,\vec{m};1) \] 
and converting once more the values at \( x=-1 \) of the $\Hpl$'s into 
values at \( x=1 \) by means of \Eq{eq:minsign}. 
\end{itemize} 
\par 
The set of relations obtained in that way is highly redundant; it has been 
checked explicitly that they generate the table of the \( w=4 \) definite 
integrals given in Appendix B of the second reference of \cite{KMR}. 
It has not yet been investigated whether they are sufficient, by 
themselves, to generate also the tables of higher weights obtained 
in \cite{HS}.
 
\par
Another powerful method to obtain the values at $x = 1/2$ when there are no 
negative indices is by considering the transformation $x \rightarrow 
z/(1+z)$, which corresponds to a suitable combination of the 
transformations $x \to 1/x$ and $x \to (1-x)$. Using the same techniques as 
in the section on related arguments, all these objects are directly 
expressed in terms of $\Hpl$-functions in $x=1$. Such expressions can then 
be used in reverse to obtain the numerical values of the `independent 
constants' that occur in the expressions for the $\Hpl$-functions at $x=1$. 
As an example we have
\begin{eqnarray}
	\Li_3\biggl(\frac{1}{2}\biggr) & = & \frac{7}{8}\zeta_3
       - \frac{1}{2}\zeta_2\ln(2) + \frac{1}{6} \ln^3(2)
\end{eqnarray}
which is of course well known. We have also
\begin{eqnarray}
	\Hpl_{2,1}\biggl(\frac{1}{2}\biggr) & = & \frac{1}{8}\zeta_3
						 - \frac{1}{6} \ln^3(2)
\end{eqnarray}
Both relations provide a power series for
the evaluation of $\zeta_3$. 
The method gives also an expression of 
$s_6 = S_{-5,-1}(\infty)$ in terms of $\Hpl_{5,1}(1/2)$, $\Hpl_6(1/2)$ and 
combinations of constants of a lower weight. Similar dependencies can be 
derived for the higher weight constants.
\par 
Let us finish with a few remarks on the numerical evaluation of the 
$\Hpl$'s for arbitrary values of \( x \). According to the discussion of 
Section 3, it is sufficient to restrict ourselves to the $\Hpl$'s either of 
the reduced set or of a 
minimal set, as all the others can be obtained from them as suitable 
combinations. The $\Hpl$'s of such a set have 
no trailing indices equal to \( 0 \), so that they can be expanded in series 
of \( x \) around \( x=0 \). For small values of \( x \) the series will 
be rapidly convergent, but the convergence will slow down approaching 
the cuts at \( x=\pm1 \). But for \( x \) approaching \( 1 \) we can use 
the transformation \Eq{eq:rel2}, so that the corresponding 
\( t = (1-x)/(1+x) \) will fall in the region near \( 0 \) and the 
expansion in \( t \) will be rapidly converging. 
\par 
More exactly, the equation 
\begin{equation}
 r = \frac{1-r}{1+r}
\end{equation}
has the two solutions \( r = - 1 - \sqrt{2} \) and \( r = - 1 + \sqrt{2} \). 
Therefore, we can use the expansion around $x= 0$ in the interval 
$ -(\sqrt{2}-1) < x < \sqrt{2}-1 $, where $|x| < \sqrt{2}-1 < 1/2$, 
switching for $ \sqrt{2}-1 < x < \sqrt{2}+1 $ to $t = (1-x)/(1+x)$, 
which corresponds to $|t| < \sqrt{2}-1 <1/2$. For greater values of 
\( x \), one can use the \( x \to 1/x \) identities. 
For large negative values of \( x \), {\it i.e.} 
\( x < 1-\sqrt{2} \), one can flip the sign of \( x \) with \Eq{eq:minsign} 
and then proceed as above. 
\par
In practice the transformation of \Eq{eq:rel2} can lead to a 
large number of functions to be evaluated and hence it may be more 
profitable to apply this transformation only for values of $x$ that are 
much closer to one. If, on the other hand, nearly all $\Hpl$-functions of a 
given weight have to be evaluated for some value of $x$ one can use the 
turnover value of $\sqrt{2}-1$ in a rather profitable way.

\par
The values in $x=1$ require some extra attention. These are actually needed 
rather frequently and hence there exists some literature on them. From 
\Eq{eq:recsum} it should be clear that an $\Hpl$-function in $x=1$ can be 
expressed in terms of either $S$-sums or $Z$-sums in infinity. Hence much 
information can be found in \cite{BBB}, \cite{BBBL} and the papers they 
refer to. Ref.~\cite{HS} gives a different method to evaluate these sums.
Recently this method has been used by one of us (J.V.) to obtain all such 
sums up to weight 9 (see also footnote 2). For only nonnegative indices 
results have been 
obtained up to weight 11~\cite{DBprivate}. When the first index of the 
$\Hpl$-function (or the $S$-sum) is one, the value in $x=1$ (or the sum in 
infinity) will be divergent. Yet we have to consider these objects. As 
mentioned in the section on the algebra this can be done consistently only 
in terms of the sums. Hence the safest method is to rewrite the 
$\Hpl$-functions in $x=1$ immediately in terms of either $S$-sums or 
$Z$-sums. In the case that the weights are low enough, these can then be 
rewritten in terms of a limited set of `fundamental constants'.


\section{Mellin transforms}

At times one may need the Mellin transform of the Harmonic polylogarithms. 
In ref~\cita{HS} a method is given to evaluate such transforms for a class 
of functions which is more or less the class of $\Hpl$-functions. There is 
however one complication with Mellin transforms. Divergencies at $x=1$ 
must be extracted. This is because the Mellin transform is defined by
\begin{eqnarray}
\labbel{eq:mellin}
 M(f(x),N) & = & \int_0^1dx\ x^N f(x) \nonumber \\
 M(\frac{f(x)}{(1\minus x)_+},N) & = & \int_0^1dx\ \frac{x^N 
       f(x)-f(1)}{1\minus x} \nonumber \\
 M(\frac{f(x)\ln^p(1\minus x)}{(1\minus x)_+},N) & = &
   \int_0^1dx\ \frac{\left(x^N 
       f(x)-f(1)\right)\ln^p(1\minus x)}{1\minus x}
\end{eqnarray}
in which the function $f$ is supposed to be finite for $x=1$ when the 
factor $1/(1\minus x)_+$ is present. Hence we have to pay attention to 
the powers of $\ln(1\minus x)$. They can be isolated with 
\Eq{eq:ln(1-x)}. After this extraction the remaining 
$\Hpl$-functions are finite in $x=1$.

At this point we can attack the Mellin transforms. It is easy to obtain 
the lowest weight results:
\begin{eqnarray}
 \int_0^1dx x^n \Hpl(0;x) & = & -\frac{1}{(n+1)^2}
  \nonumber \\
 \int_0^1dx x^n \Hpl(1;x) & = & \frac{S_1(n+1)}{n+1}
  \nonumber \\
 \int_0^1dx x^n \Hpl(-1;x) & = & \sign(n)\frac{S_{-1}(n+1)}{n+1}
  +\frac{\ln(2)}{n+1}\left(1+\sign(n)\right)
\end{eqnarray}
in which we have used that $\Hpl(-1;1) = -S_{-1}(\infty) = \ln(2)$.
The higher weight results can be obtained by recursion. 
Like in ref~\cita{HS} this is done by partial integration. We 
also exchange the sums immediately after each step so that we may do one of 
them immediately. The result is:
\begin{eqnarray}
\labbel{eq:recmellin}
 \int_0^1 dx\sum_{i=n}^\infty \sigma^i x^i \Hpl_{0,\vec{m}}(x) 
   \frac{S_{\vec{p}}(i\plus 1)}{(i\plus 1)^k} & = &
   \sigma \Hpl_{0,\vec{m}}(1) \left(
     S_{\sigma(k\plus 1),\vec{p}}(\infty)
     -S_{\sigma(k\plus 1),\vec{p}}(n)\right)
    \nonumber \\ &&
   -\int_0^1 dx\sum_{i=n}^\infty \sigma^i x^i \Hpl_{\vec{m}}(x) 
   \frac{S_{\vec{p}}(i\plus 1)}{(i\plus 1)^{k\plus 1}}
  \\
 \int_0^1 dx\sum_{i=n}^\infty \sigma^i x^i \Hpl_{1,\vec{m}}(x) 
   \frac{S_{\vec{p}}(i\plus 1)}{(i\plus 1)^k} & = &
   \sigma \Hpl_{1,\vec{m}}(1) \left(
     S_{\sigma(k\plus 1),\vec{p}}(\infty)
     -S_{\sigma(k\plus 1),\vec{p}}(n)\right)
    \nonumber \\ &&
   -\int_0^1 dx\sum_{i=n}^\infty x^i \Hpl_{\vec{m}}(x)\Bigl( 
   \sigma S_{\sigma(k\plus 1),\vec{p}}(i\plus 1)
    \nonumber \\ &&
   -\sigma^i \frac{S_{\vec{p}}(i\plus 1)}{(i\plus 1)^{k\plus 1}}
   -\sigma S_{\sigma(k\plus 1),\vec{p}}(n)
   \Bigr)
  \\
 \int_0^1 dx\sum_{i=n}^\infty \sigma^i x^i \Hpl_{\minus 1,\vec{m}}(x) 
   \frac{S_{\vec{p}}(i\plus 1)}{(i\plus 1)^k} & = &
   \sigma \Hpl_{\minus 1,\vec{m}}(1) \left(
     S_{\sigma(k\plus 1),\vec{p}}(\infty)
     -S_{\sigma(k\plus 1),\vec{p}}(n)\right)
    \nonumber \\ &&
   -\int_0^1 dx\sum_{i=n}^\infty x^i \Hpl_{\vec{m}}(x)\Bigl( 
   \sigma\sign(i) S_{\minus\sigma(k\plus 1),\vec{p}}(i\plus 1)
    \nonumber \\ &&
   +\sigma^i \frac{S_{\vec{p}}(i\plus 1)}{(i\plus 1)^{k\plus 1}}
   -\sigma\sign(i) S_{\minus\sigma(k\plus 1),\vec{p}}(n)
   \Bigr)
\end{eqnarray}
The variable $\sigma$ is either $1$ or $-1$. This leaves only the 
evaluation of the $\Hpl$-functions in $x=1$. These values do not have to be 
finite. Only the $\Hpl$-functions that are used in the subtraction in 
\Eq{eq:mellin} are finite. This causes no problems provided the 
divergencies are regularized in the representation in terms of $S$-sums as 
explained before. 

As an example we show here a nontrivial Mellin transform:
\begin{eqnarray}M\Biggl(\frac{H_{1,\minus 2,1,0}(x)}{1-x},N\Biggr) & = &
 S_{1,\minus 2,\minus 1,2}(N)
 -2S_{1,\minus 2,\minus 3}(N)
 +2S_{1,5}(N)
  \nonumber \\ &&
 -\frac{1}{2}S_{1,\minus 2}(N)\Biggl(\zeta_2\ln(2)+\zeta_3\Biggr)
 +S_{1,2}(N)\Biggl(\frac{1}{2}\zeta_2\ln(2)-\zeta_3\Biggr)
  \nonumber \\ &&
 +S_{1,1}(N)\Biggl(4\Li_4\biggl(\frac{1}{2}\biggr)+\frac{1}{6}\ln^4(2)
  -\zeta_2\ln^2(2)-\frac{13}{40}\zeta_2^2\Biggr)
  \nonumber \\ &&
 +S_1(N)\Biggl(\frac{9}{2}\zeta_2\zeta_3-\frac{83}{8}\zeta_5\Biggr)
 -\frac{1}{24}\zeta_2\ln^4(2)
 +\frac{7}{16}\zeta_2\zeta_3\ln(2)
  \nonumber \\ &&
 -\zeta_2\Li_4\biggl(\frac{1}{2}\biggr)
 +\frac{1}{4}\zeta_2^2\ln^2(2)
 +\frac{447}{840}\zeta_2^3
 -\frac{157}{32}\zeta_3^2
 +\frac{7}{2}S_{\minus 5,\minus 1}(\infty)
\end{eqnarray}
The sum in the last term is irreducible.

In the case that the weight of the terms too large (currently larger than 
9) it becomes 
rather hard to obtain the values for the $\Hpl$-functions in $x=1$ or 
alternatively for the $S$-sums in infinity. Because the algebra for the 
$\Hpl$-functions in $x=1$ is different from the algebra for the 
$\Hpl$-functions for general values of $x$ there may be large numbers of 
$\Hpl$-functions left that each are divergent at $x=1$. The reason is that 
some algebraic work is done first with the general algebraic rules and has 
to be `undone' with the rules for $x=1$. 
The relations that 
make the divergences cancel may not be easy to find. One can still obtain 
numerical results however. 

If one is faced with higher weights one may proceed as follows. The 
$\Hpl$-functions in $x=1$ are first expressed in terms of $S$-sums in 
infinity. Then the shuffle algebra for the $S$-sums is used to extract the 
divergencies in a way that is similar to how this is done for the powers of 
$\ln(1-x)$ for the $\Hpl$-functions. Because the divergences have to cancel 
each other, all divergent terms should disappear, even though we may not 
have the algebraic methods to prove this for the case at hand. The 
remaining finite expression can in principle be evaluated numerically.

Inverse Mellin transforms are now relatively easy. As pointed out in 
ref~\cita{HS} each $S$-sum has a single most complicated original function 
in terms of $\Hpl$-functions in which we can define `most complicated' by 
function with largest weight or in the case of identical weights the 
largest number of nonzero indices. And actually one can obtain the relation 
between the $S$-sum of which one needs the inverse Mellin transform and 
this most complicated $\Hpl$-function from the recursion relations in 
\Eq{eq:recmellin}. Hence the algorithm is clear:
\begin{itemize}
\item Locate the most complicated $S$-sum(s).
\item Construct the corresponding $\Hpl$-function(s) in $x$-space.
\item Add it and subtract it.
\item Make the Mellin transform of the subtracted version.
 This will cancel the original $S$-sum.
\item Repeat the above steps until there are no more $S$-sums remaining.
\item Multiply the remaining constant terms by $\delta(1\minus x)$.
\end{itemize}
This algorithm will properly terminate. It has only one problem: Some 
Mellin transforms have a factor $\sign(N)$ and some don't. What if we take 
an $S$-sum which should have a factor $\sign(N)$ but we omit it? Here we 
have to realize that the inverse Mellin transform is to be constructed from 
either all even or from all odd moments only. Hence we have to specify 
whether $N$ is even or odd. This will give a value to $\sign(N)$. Hence the 
only thing that remains is to give the relation between an $S$-sum and the 
most complicated $\Hpl$-function that contributes to it.
\begin{itemize}
\item If the number of negative indices is odd, there will be a factor 
$1/(1\plus x)$, otherwise there will be a factor $1/(1\minus x)_+$.
\item Next copy the index field to the $\Hpl$-function.
\item Working from the rightmost index to the left, each index will get a 
sign that is the combination of its old sign and the signs of all indices 
to the left of it.
\item There will be an additional overall sign on the term which is the sign 
of the last index.
\item There will be an additional overall sign on the term which is 
$\sign(w-1-d)$ in which $w$ is the weight of the $S$-sum and $d$ its 
depth (which is the number of nonzero indices).
\item Each negative index in the current configuration will give a minus 
sign to the term.
\end{itemize}
We will give two examples of weight 7 functions. First and example that 
involves subtractions with $\ln^2(1\minus x)$ in the Mellin transform:
\begin{eqnarray}
 S_{1,1,2,1,2}(N) & \rightarrow &
  \frac{1}{1\minus x}\Biggl(
       -\Hpl_{1,1,2,1,0}(x)
       -\frac{1}{2} \Hpl_{1,1}(x) \zeta_2^2
          \Biggr)
  \nn
  +\delta(1\minus x)\Biggl(
       - \frac{3}{2}\zeta_2\zeta_5 - \frac{7}{5}\zeta_2^2\zeta_3 + 17\zeta_7
   \Biggr)
\end{eqnarray}
In this case there is no difference for even values of $N$ and for odd 
values of $N$.
However the next example is different. For even values of $N$ we have
\begin{eqnarray}
 S_{\minus 1,1,\minus 2,1,2}(N) & \rightarrow &
  \frac{1}{1\minus x}\Biggl(
       - \Hpl_{-1,-1,2,1,0}(x)
       + 2\Hpl_{-1,-1,0,0,0,0}(x)
       - 2\Hpl_{0,0,0,0,0,0}(x)
 \nn
    \ \ \   + \frac{1}{2}\Hpl_{-1,-1}(x) \zeta_2^2
       - \Hpl_{0,0}(x) \zeta_2^2
 \nn
    \ \ \   + \Hpl_{-1}(x) ( \frac{1}{16}\zeta_2\zeta_3
     + \zeta_2^2\ln(2) + \frac{67}{64}\zeta_5 )
 \nn
    \ \ \   + \Hpl_{0}(x) ( - \frac{1}{8}\zeta_2\zeta_3
     - 2\zeta_2^2\ln(2) - \frac{67}{32}\zeta_5 )
 \nn
    \ \ \   - \frac{1}{8}\zeta_2\zeta_3\ln(2)
    - \frac{3}{2}\zeta_2^2\ln^2(2)
    - \frac{307}{560}\zeta_2^3
       + \frac{157}{128}\zeta_3^2
       + \frac{21}{64}\zeta_5\ln(2)
       - \frac{5}{4}\sigma_6
  \Biggr)
 \nn
  +\frac{1}{1\plus x}\Biggl(
       (\Hpl_{1}(x)+2\Hpl_{0}(x)) 
    ( \frac{1}{16}\zeta_2\zeta_3 - \frac{53}{64}\zeta_5 )
 \nn
    \ \ \   - \frac{61}{560}\zeta_2^3
    + \frac{35}{128}\zeta_3^2
       + \frac{93}{64}\zeta_5\ln(2)
       - \frac{3}{4}\sigma_6
  \ \Biggr)
 \nn
 +\delta(1\minus x)\Biggl(
       - \frac{1}{16}\zeta_2\zeta_3\ln^2(2)
       - \frac{957}{224}\zeta_2\zeta_5
       + \frac{1}{120}\zeta_2\ln^5(2)
       - \zeta_2\Li_5(\frac{1}{2})
 \nn
	\ \ \ - \frac{93}{140}\zeta_2^2\zeta_3
       - \frac{1}{12}\zeta_2^2\ln^3(2)
       - \frac{29}{280}\zeta_2^3\ln(2)
       - \frac{1355}{896}\zeta_3^2\ln(2)
 \nn
    \ \ \ - \frac{197}{64}\zeta_5\ln^2(2)
       + \frac{37215}{3584}\zeta_7
       + \frac{19}{28}\ln(2)\sigma_6
       - \frac{10}{7}\sigma_{7,a}
       + \frac{29}{14}\sigma_{7,b}\ \Biggr)
\end{eqnarray}
in which
\begin{eqnarray}
 \sigma_6 & = & S_{-5,-1}(\infty)
  \nonumber \\
 \sigma_{7,a} & = & S_{-5,1,1}(\infty)
  \nonumber \\
 \sigma_{7,b} & = & S_{5,-1,-1}(\infty)
\end{eqnarray}
In the case of odd $N$ the terms with $1/(1\plus x)$ change sign. 
As one can see these formulae can become rather involved, even though the 
number of terms is rather small compared to the number of functions that 
exist in $x$-space for this weight.

In the case that sums of a higher weight are considered one may not be able 
to substitute the values of the $\Hpl$-functions at $x=1$. The same 
considerations as for the Mellin transforms can be used to obtain 
an answer that can at least be evaluated numerically. In general the 
formulae will of course be much lengthier.


\vskip 2 truecm 
{\large\bf Acknowledgements.}
One of the authors (E.R.) wants to thank the Alexander von Humboldt 
Stiftung for the generous support of his stay at Karlsruhe.
The other author (J.V.) would like to thank the Programa "Catedra" of the 
Fundacion BBV for support during the part of this work which was done at 
the Universidad Aut\'onoma of Madrid.

\noindent We like to thank S. Moch for discussions.
\vskip 2 truecm 
\def\NC{{\sl Nuovo Cimento }\ } 
\def\NP{{\sl Nuc. Phys. }\ } 
\def\PL{{\sl Phys. Lett .}\ } 
\def\PR{{\sl Phys. Rev. }\ } 
\def\PRL{{\sl Phys. Rev. Lett. }\ } 

\end{document}